\documentclass[a4paper,10pt]{article}
\usepackage{graphics}

\begin{document}
\title{{\bf{Comment to Manipulating the polarization of laser beams by an external magnetic field,
arXiv:physics/0603175v1 21 Mar 2006}}}
\author{R. Battesti\footnote{Laboratoire National des Champs Magn\'{e}tiques Puls\'{e}s,
CNRS/INSA/UPS, UMR 5147, 31432 Toulouse, France.} , B. Pinto Da
Souza\footnotemark[2],  C. Rizzo\footnotemark[2], C.
Robilliard\footnotemark[2], J. Vigu\'{e}\footnote{IRSAMC,
Laboratoire Collisions Agr\'{e}gats R\'{e}activit\'{e}, UPS/CNRS,
UMR 5589, 31062 Toulouse, France.}
}

\maketitle

The basic assumption of the authors of manuscript
arXiv:physics/0603175v1 21 Mar 2006 is that a laser wave differs
from a plane wave because of photon coherence. This is not true
and therefore their conclusion is also wrong.

The authors claim that a laser wave is a coherent state of photons
with momentum $\bf{k}$ and polarization $\lambda$, while a plane
wave is not. Actually, a plane wave is a coherent state which
coherence length $l_c$ is infinite, where $l_c=c/\Delta \nu$, $c$
being the light velocity and $\Delta \nu$ the spectral width of
the wave packet. Obviously $\Delta \nu$ is zero for a plane wave.

Thus the $\kappa^{(1)}(\omega)$ of equation 10 does not represent
the absorption rate for a plane wave but simply the absorption
rate for a single incident photon. Therefore equation 10 simply
states that the total absorption rate is proportional to the
number of photons, which is trivial. The enhancement of the
absorption rate of a laser beam compared to that of a single
photon has thus nothing to do with quantum coherence. The same
announcement can be made for a plane wave having the same mean
number of photons.

Moreover, the authors claim that their calculations may explain
for instance the results of the PVLAS collaboration (See ref. [2]
of the original manuscript), but they do not explain how they
figured it out. Let us make the calculation: according to the
authors, the ellipticity and apparent dichroism should be enhanced
by a factor

$$\bar{N} \simeq \frac{\epsilon_0 E^2}{\hbar \omega} S l_c, $$

\noindent where $E \sim 1$ kV/m corresponds to a power of 100 mW,
$\hbar \omega \sim 2\times 10^{-19}$ J corresponds to a wavelength
of 1.06 $\mu$m, $S \sim 10^{-6}$ m$^2$ the waist of the beam and
$l_c = c/\Delta \nu \sim 10^6$ m of the coherence length of the
cavity.

As a result, the enhancement factor should be at least on the
order of $\bar{N} \sim 10^{13}$, i.e. 9 orders of magnitude larger
than what was indeed observed!

\end{document}